\documentstyle[prl,aps,epsf]{revtex}
\draft
\begin{document}
\twocolumn[\hsize\textwidth\columnwidth\hsize\csname @twocolumnfalse\endcsname

\title{Nonlinear dynamics, rectification, and phase locking for 
particles on symmetrical two-dimensional periodic substrates 
with dc and circular ac drives
}
\author{C. Reichhardt, C.J. Olson Reichhardt, and M.B.~Hastings} 
\address{ 
Center for Nonlinear Studies and Theoretical Division, 
Los Alamos National Laboratory, Los Alamos, New Mexico 87545}

\date{\today}
\maketitle
\begin{abstract}
We investigate the dynamical motion of particles on a 
two-dimensional symmetric 
periodic substrate in the presence of both
a dc drive along a symmetry direction of the periodic substrate 
and an additional circular ac drive. 
For large enough ac drives, the particle orbit encircles one or more potential 
maxima of the periodic substrate. 
In this case, when an additional increasing 
dc drive is applied in the longitudinal 
direction, the longitudinal velocity increases in a 
series of discrete steps that are
integer multiples of $a\omega/(2\pi)$, 
where $a$ is the lattice constant of the substrate. 
Fractional steps 
can  also occur. These integer and fractional steps correspond to 
distinct stable dynamical orbits. 
A number of these phases also show a 
rectification in the positive or negative 
transverse direction where a non-zero 
transverse velocity occurs in the absence of a dc transverse drive. 
We map out the phase diagrams of the regions of  
rectification as a function of ac amplitude, and find a series of tongues. 
Most of the features, including the steps in the longitudinal 
velocity and the transverse rectification,  
can be captured with a simple toy
model and by arguments from nonlinear maps. 
We have also investigated the effects
of thermal disorder and incommensuration 
on the rectification phenomena, and find that for increasing disorder,   
the rectification regions are 
gradually smeared and the longitudinal velocity steps
are no longer flat but show a linearly increasing velocity. 

\end{abstract}

\vskip2pc]
\narrowtext

\section{Introduction}

Recently there has been a growing interest in  
nonequilibrium systems that show a
rectification or ratchet effect, typically  
for a particle moving in some form 
of asymmetric potential \cite{Reviews1}.  In these systems, 
a net dc drift in one direction can occur 
even though only an
ac drive or ac flashing of the potential is applied. 
Such ratchet phenomena have been examined in a variety
of systems, including
biological motors \cite{Rachet2}, 
colloidal particles moving through asymmetric potentials 
\cite{Rachet2,Det3}, atom transport in optical lattices 
\cite{Atom4}, charge transport in quantum dot systems \cite{Linke5},
transport of granular particles \cite{Farkas6}, and
vortices in
superconductors and SQUIDs \cite{FluxRatchet7,Hanggi8}.
In most of these systems
there is some form of underlying asymmetric substrate 
potential which is responsible for the symmetry breaking that 
gives rise to the rectification.    
Additionally, most of the systems studied so far have 
one-dimensional (1D) or
effectively 1D geometries. 

For 2D systems, it is possible to break the symmetry 
of the system without introducing an asymmetric substrate. 
One example of rectification in 2D is
the motion of biomolecules or
polymers through periodic arrays of posts
\cite{Austin9,Viovy10}. Here
the particles are driven in alternating directions
by an electric field.  Another 
approach to 2D rectification is to drive particles through 
a periodic array at various angles \cite{Grier11,Korda12,GrierN13}. 
The particle motion becomes locked to certain stable angles,
such as $45^{\circ}$ for a square array, 
even when the
external drive is applied in a different direction.    
Several theoretical studies have also considered
models of particles moving in 2D asymmetric 
potentials, leading to rectification and negative differential
conductivity \cite{Cecchi14,Reimann15}.
In a recently proposed model, 
spatiotemporal symmetry breaking is caused by the application of 
an external wave to a system with a periodic potential \cite{Zheng16}.  
In other models,
the asymmetry of quantities other
than the substrate produces a rectification
\cite{Lutz17}. 
A better understanding of
2D systems that exhibit rectification
can 
assist in the creation of
technological devices for applications such as      
the separation of different species
of colloids or biomolecules or new techniques for electrophoresis.  

The phase locking that occurs when particles are driven over
periodic substrates in the presence of an ac drive has also been
intensely studied.
This phenomenon arises
when the external ac frequency $\omega$ matches the
internally generated  frequency of the
motion of the particle over the periodic potential. 
One of the best known examples of phase locking is the
Shapiro steps observed as steps in the V(I) curves of Josephson-junction arrays
\cite{Shapiro18}.  The step widths oscillate with the
ac amplitude $A$, with the $n$th step varying as the modified Bessel
function $J_{n}(A/\omega)$. 
Shapiro step-like phase locking is also observed 
for dc and ac drives in sliding charge-density wave
systems \cite{Thorne19}, as well as vortex motion in superconductors
with periodic substrates \cite{Martinoli20,Look21,Zimanyi22}.

In typical phase-locking systems,
the ac drive is applied in the {\it same} direction as the dc drive.
Additionally, most of the well-studied phase locking 
systems can be considered as effectively 1D.
Phase locking should also occur in
2D when the ac drive is applied in a different direction 
from the dc drive; 
however, very little is known about the behavior of phase locking
in this case. 
Rectification may occur
even for motion in a symmetric potential 
if the ac drive in a 2D system breaks the symmetry,
such as a circular ac drive.   

In a recent model for vortices
in a 2D superconductor moving over a periodic potential, the 
ac drive was {\it perpendicular} to the dc drive \cite{Phase23}. 
In this case the
phase locking that occurred was distinct from Shapiro steps. 
For these perpendicular ac drives, the widths of the steps do
not oscillate with the drive amplitude,
as would be expected for Shapiro steps, 
but instead they monotonically increase as the square of the ac amplitude for
square substrates and linearly for triangular substrates.  

For elastic media moving over 
a {\it random} substrate, it is also possible to have a periodic 
velocity component in both the longitudinal and transverse directions 
due to the periodicity of the elastic media itself.
When an ac drive is applied in the same direction as the dc drive 
for random disorder,
Shapiro-like phase locking effects can again be observed, 
such as in
sliding charge density waves 
\cite{Thorne19} and vortex lattices \cite{Harris24,Kolton25,Kobubo26}.
In recent simulations and theory for the case of a perpendicular ac drive
for vortex lattices interacting with random pinning, 
a transverse phase
locking occurs \cite{Dominguez27}.
In 2D, it is possible to apply {\it two} ac drives 
which are perpendicular to one another such that the particle, in the
absence of a dc drive, would move in a circle.     
The behavior of the system in this case has been largely unexplored.

In this work we study the motion of overdamped particles moving over a 
two-dimensional symmetric periodic
substrate where there are two perpendicular ac drives and an additional
dc drive that is applied along a symmetry direction of the
substrate. 
In Section II we outline our model and 
the simulation technique. 
In Section III, 
we show that for small ac amplitudes
and low dc drives, the particle moves in a 
circular orbit 
near the potential minimum. As a function of the applied dc driving
force $f_{dc}$ in the longitudinal direction, 
there is a depinning threshold for the particle motion.
For increasing drive beyond the threshold, the longitudinal velocity
$V_{x}$ increases
in a series of steps; however, there
is no rectification and  the transverse velocity $V_{y}$ is zero.  
For increasing ac amplitude
of fixed frequency 
and no dc drive, the circular particle orbit increases in radius, and 
there are
a series of stable orbits which are commensurate with the
periodicity of the substrate. 
In Section IV, we illustrate that when a dc drive is applied
for ac amplitudes such that
the particle orbit encircles one 
potential maximum, 
the longitudinal velocity again increases in a series of 
prominent integer steps. 
Between these integer steps, there are a 
series of smaller fractional steps with a structure 
similar to a Devil's staircase. 
In Section V, we show that for the same range of ac amplitude where 
the particle orbit encircles one
maximum, there are also distinct regions where a transverse rectification
occurs. The rectification phases are centered
between the integer steps in the longitudinal velocity.
The maximum velocity in the rectification
regions is $a\omega/(2\pi)$, and we observe smaller fractional
rectification steps as well. 
The rectification can be in either the positive or negative direction.
We map out the phase diagram of the rectification phases as
a function of ac amplitude and dc drive, and show
that it consists of a series of tongues. 
We find in general that as the ac amplitude increases,
the number of regions that show a rectification also increases.
In Section VI we examine the effects of disorder. For increasing
thermal disorder, the phases begin to smear; however, regions
of rectification persist up to high temperatures.
We also consider the effects of particle-particle interactions
when multiple particles move through the arrays. In this case,
incommensuration effects produce a partial smearing of the phases. 
In Section VII we examine a simple model that captures most of the features
of the system, including integer and fractional steps in the
longitudinal velocity as well as steps in the transverse velocity that
correspond to positive or negative rectification.
In Section VIII we discuss some experimental systems in which the 
phase locking and ratchet effects might be observed, including
colloids moving through periodic traps, biomolecules driven through
arrays of posts, vortices in superconductors with periodic pinning arrays,
and classical electron motion in antidot arrays.
A shorter version of portions of the work presented here has been
previously published \cite{Ratchetnew28}.

\section{Simulation}

We consider an overdamped particle moving in two dimensions 
and interacting 
with an underlying square periodic substrate, 
where we use periodic boundary conductions
in the $x$ and $y$ directions.
The equation of motion for a particle $i$ is
\begin{equation}
{\bf f}_{i} =  
{\bf f}_{s} + {\bf f}_{dc} + {\bf f}_{ac} = 
\eta{\bf v}_{i}, 
\end{equation}
where  the damping constant
$\eta$ is set to unity. 
The substrate force, ${\bf f}_{s}$, 
arises from a square array with period $a$
of repulsive sites, each of which has 
force
${\bf f}^j_{s} = -\nabla U(r)$.
To model specific physical systems we consider 
$U(r) = \ln(r)$, corresponding to a thin film superconductor
with a periodic array of pinning sites, where each pinning site 
captures one vortex and an additional vortex sits in the 
interstitial region between pinning sites. 
This interstitial vortex interacts with a
square periodic substrate created by the pinned vortices. 
We have also considered
$U(r)=1/r$ and $U(r)=e^{-r\kappa}/r$,  
which could model one mobile particle in 
an array of trapped colloids or ions. 
We have considered systems of different sizes
and find that for most of the results
presented here, $8a\times8a$ is adequate.
The dc drive ${\bf f}_{dc}$ is applied
along the symmetry axis of the  
substrate array, in the $x$ or longitudinal direction. 
The ac drive has two components, given by

\begin{figure}
\center{
\epsfxsize=3.5in
\epsfbox{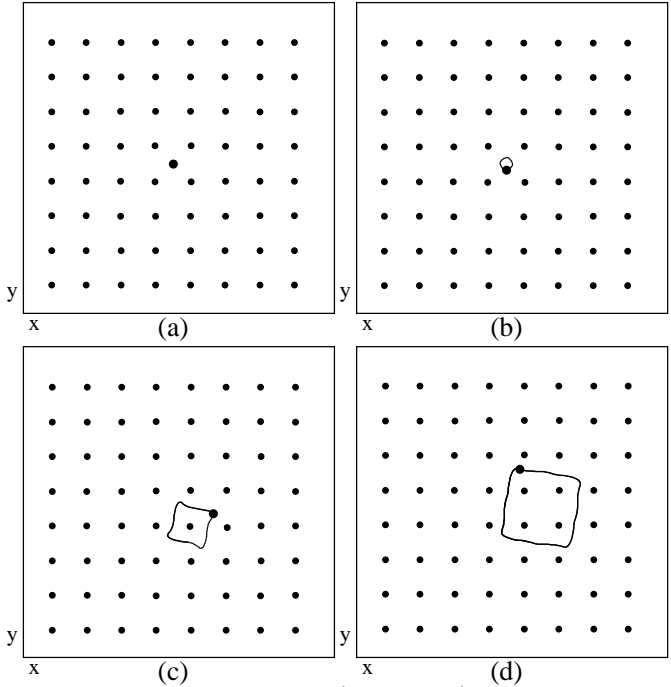}}
\caption{Potential maxima (black dots), driven particle (large dot),
and particle trajectory (black line) for ac amplitude
$A=$ (a) 0.0, (b) $0.15$,
(c) $0.3$, and (d) $0.45$. 
}
\end{figure}

\noindent
\begin{equation} 
{\bf f}_{ac} = A\sin(\omega_{A}t){\hat {\bf x}} - 
B\cos(\omega_{B}t){\hat {\bf y}}.
\end{equation}
In all of our results 
there is {\it no} 
dc driving component in the $y$ or transverse direction. 
We fix $w_{A}/w_{B} = 1.0$ and $A = B$. 
As an initial condition,
we place the particle close to the center of a plaquette. 
For different initial placements, the results are identical.
In a single simulation the dc drive ${\bf f}_{dc}$ is increased  
from $0$ to $2.0$
in increments of $0.0001$, where 
$3\times10^5$ time steps are spent at each increment
to ensure a steady state.
We measure the particle trajectories and velocities
in the longitudinal   
$V_{x} = \sum_{i}^{N_{v}} {\hat {\bf x}}\cdot {\bf v}_{i}$ 
and transverse direction
$V_{y} = \sum_{i}^{N_{v}}{\hat {\bf y}}\cdot {\bf v}_{i}$. 
We have also considered the cases $A \neq B$ and 
$\omega_{A} \neq \omega_{B}$, as well as the addition of a phase offset
and driving with more complicated ac forms. These introduce
a considerable number of new behaviors not found for the strictly circular 
case, and have been detailed in a separate publication \cite{EllipticalR29}. 

\section{Commensurate Orbits and Depinning}

In Fig.~1 we show the locations of the 
substrate potential maxima
of the form $\ln(r)$ and the
trajectories or orbits of the mobile particle for 
fixed $f_{dc} = 0.0$ 
and ac amplitudes of $A=$ (a) $0.0$, (b) $0.15$,
(c) $0.3$ and (d) $0.45$. 
For $A = 0$ [Fig.~1(a)], the particle is
stationary and is located at the center of a plaquette at the potential
minimum. 
For $0.0 < A < 0.25$, the particle moves 

\begin{figure}
\center{
\epsfxsize=3.5in
\epsfbox{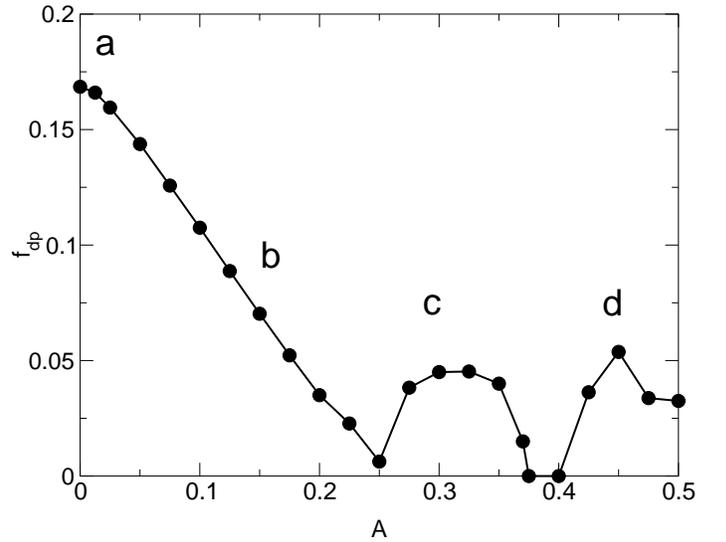}}
\caption{
Depinning force $f_{dp}$ vs ac amplitude $A$. The letters
a, b, c, and d correspond to the amplitudes
at which the orbits shown in 
Fig.~1(a-d) occur. 
}
\end{figure}

\noindent
in a circular
orbit around the center of the plaquette,
as seen in Fig.~1(b). 
The radius of the orbit increases with $A$, and the orbit
becomes increasingly square as $A$ approaches $0.25$, 
reflecting the square symmetry of the 
caging potential in the plaquette.  
For $0.25 < A < 0.375$, the 
radius of the orbit is large enough that, during a single periodic cycle, 
the particle encircles one potential maximum,  
as shown in Fig.~1(c). 
For $0.4 < A < 0.5$, the particle moves in a stable
orbit that encircles four potential maxima,
as in Fig.~1(d). Between the regions where four and one maxima are encircled,
for $0.375 < A < 0.4$, the  
orbits are unstable and the particle is no longer localized
but undergoes diffusion. 
Stable orbits also occur for higher
ac amplitudes where $9$ and $25$ maxima are encircled, 
with regions of delocalized particle motion falling between
the ac amplitudes that produce stable orbits.
A similar 
phenomena of commensurate orbits for particles undergoing circular or
cyclotron motion in a periodic array of scatterers also occurs
in electron pinball models \cite{Weiss30}. Another
similar system is  
the vortex pinball model, where stable
orbits occur for vortices in superconductors with periodic pinning
arrays when the density of the magnetic field is such that there are
twice as many vortices as pinning sites \cite{vortexpinball31}.

In Fig.~2 we plot the depinning threshold 
$f_{dp}$ vs ac amplitude $A$ for the
system in Fig.~1 under the application of a dc force.
The depinning threshold decreases continuously 
with increasing $A$ for $0.0 < A < 0.25$ 
when the particle is circling inside a single plaquette, 
as illustrated in Fig.~1(b).  
For $A > 0.25$, the depinning threshold increases with $A$
and reaches a local maximum at $A = 0.3$, corresponding
to the center of the range of ac amplitudes 
over which the stable orbit encircles one potential maximum
[Fig.~1(c)]. 
The depinning threshold drops to zero for $ 0.375 < A < 0.4$, 
when the particle is delocalized. 
For 

\begin{figure}
\center{
\epsfxsize=3.5in
\epsfbox{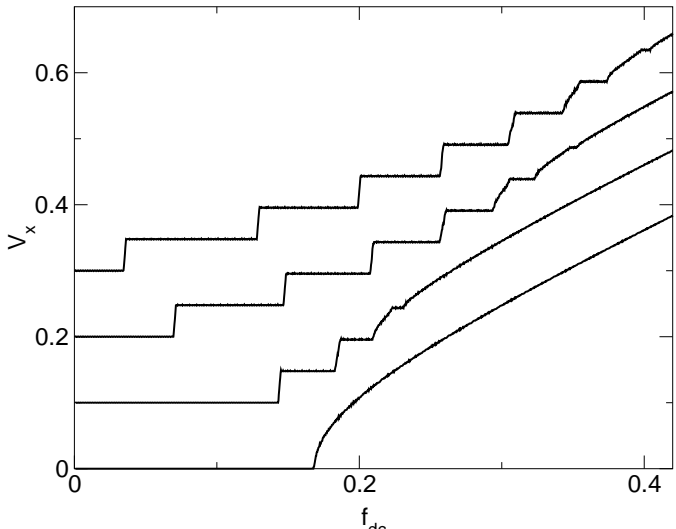}}
\caption{
The velocity in the $x$-direction, $V_{x}$, vs $f_{dc}$ for $A = 0.0$, 
0.05, 0.15,
and $0.2$ from bottom to top. The curves have been systematically 
shifted in $y$
for clarity.  
}
\end{figure}

\noindent
$A > 0.4$, $f_{dp}$ again increases with $A$ to a local
maximum at $A=0.45$ at the center of the region where the particle
orbit encircles four potential maxima.
We also find
non-zero depinning thresholds for higher 
values of $A$ at which $9$ and $25$ potential maxima
are encircled by the orbit in a single period. 

\section{Phase Locking For Low AC Amplitudes} 

We now consider the phase locking phenomena for 
low ac amplitudes 
$0 < A < 0.25$, when the particle moves in the interstitial region
between the potential maxima as in Fig.~1(b). 
In Fig.~3 we plot $V_{x}$ vs $f_{dc}$
for increasing ac amplitudes $A = 0.0$, 0.5, 0.15, and $0.2$,
showing that a series of steps
occur which increase in width with increasing $A$ from zero at $A=0$.
The depinning threshold decreases
with increasing $A$. 
The height of the $n$th step is $V_x=na\omega/(2\pi)$, 
and as $A$ increases, higher order steps can be resolved.  
If the ac drive were applied in the
$x$-direction only, Shapiro-type steps would 
occur with $V_x=na\omega/(2\pi)$ on the $n$th step. 
For Shapiro type phase locking, the 
velocity vs force curve at the beginning and end of each step
would be continuous; however, the edges
of the steps in Fig.~3 are extremely sharp. 
The average velocity
component in the $y$-direction is strictly zero for all values of
$A < 0.25$. For varied $\omega$ the location of the steps shifts; 
however, the same general behavior occurs.

The quantization of the step height is a result of
the periodicity of the drive.  
Although the particle has translated by $n$ unit cells in the
$x$-direction 
after a single
period of the drive, 
the particle is in the {\it same location} within
its unit cell as it was at the start of the period.  
Thus, up to this translation by $n$ cells, the orbit
is 

\begin{figure}
\center{
\epsfxsize=3.5in
\epsfbox{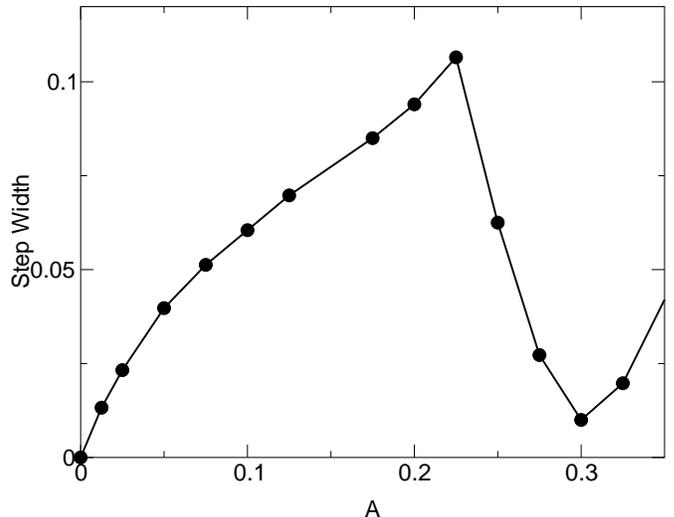}}
\caption{ 
Width of the the first phase locking step
$n=1$ vs $A$ from the curves shown in Fig.~3. 
}
\end{figure}

\noindent
periodic with period $2\pi/\omega$. 
The particle therefore moves
a distance of $na$ in a time of $2\pi/\omega$, giving a velocity of
$V_x=na\omega/(2\pi)$.

In Fig.~4 we plot the width $W$ of the $n=1$ step vs $A$ for
$0.0 < A < 0.36$. 
The width increases with $A$ for $A < 0.225$,
an ac amplitude just below the transition 
at which the particle orbit changes
from encircling zero to encircling one potential maximum
at zero dc drive.
$W$ then decreases with increasing $A$, 
reaching a minimum at $A = 0.3$, which 
corresponds to the peak in the depinning threshold shown in Fig.~2. 
For Shapiro step-type locking, the higher order step widths would
fit to a Bessel function as a function of ac amplitude.  Here,
although the step width $W$ does show an oscillatory behavior similar
to that of Shapiro steps, $W$ does not fit well to a Bessel function form,
particularly due to the sharp cusp at $A = 0.225$.  

In Fig.~5(a,b) we illustrate the particle trajectories 
along the first and second steps 
in $V_{x}$
for fixed $A = 0.15$, 
and in Fig.~5(c) we show an orbit for a non-step region.     
Along the $n=1$ step (Fig.~5(a)), the particle 
performs a loop at the center of each plaquette and its
motion is perfectly regular. 
For the $n=2$ step (Fig.~5(b)), the
regular particle orbit
has a kink or very small loop in every second plaquette.
For higher order steps, 
we find stable orbits similar to those shown
in Fig.~5(a,b), where 
for the $n$th step  
a small loop in the orbit occurs in every $n$th plaquette.  
For a typical non-step region, such as that illustrated
in Fig.~5(c), the orbits 
are disordered or chaotic, and the particle does not
follow any particular trajectory. We also find that some 
fractional steps can occur near the edges of the main steps with 
$V_x=pa\omega/(2\pi q)$, where $p$ and $q$ are integers. 
These fractional steps are much
smaller in width than the integer steps.  

The origin of the fractional steps is similar to that of the
integer steps.  
Consider a particle that has started in 

\begin{figure}
\center{
\epsfxsize=3.5in
\epsfbox{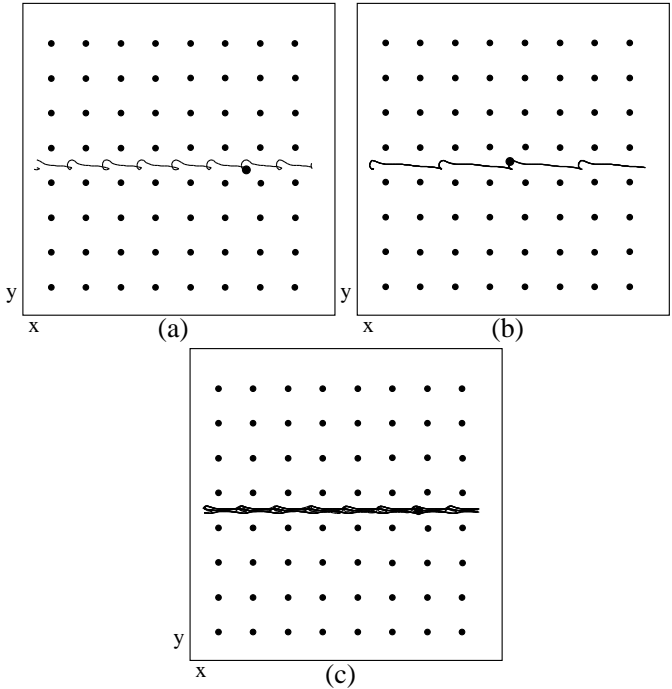}}
\caption{Potential maxima (black dots), driven particle (large dot), and
particle trajectories (black lines) for the sample in
Fig.~3 at $A=0.15$. (a) The particle orbit 
on the $n=1$ step in $V_{x}$ 
at $f_{dc} = 0.1$.
(b) The orbit for the $n=2$ step at
$f_{dc} = 0.17$. (c) The particle orbit for the non step region 
at $f_{dc} = 0.208$.} 
\end{figure}

\noindent
a given position
within some unit cell.  After a single period of the drive, the particle
may or may not have moved to another cell.  However, after a 
single period, it is in a {\it different} 
position within the unit cell than that which it 
occupied at the start.  Only after $q$ periods elapse does the particle
return to the same position in the unit cell.  Thus, up to a translation by
some number of unit cells,
the orbit is periodic with period $2\pi q/\omega$.
If the particle translates by $p$ cells in this time, we 
obtain a velocity
of $V_x=pa\omega/(2\pi q)$.

\section{Phase Locking and Rectification For Higher ac Amplitudes}

We next turn to the phase locking for ac amplitudes $0.225 < A < 0.4$, 
where the particle orbit encircles one potential maximum as shown in 
Fig.~1(c). In Fig.~6 we plot $V_{x}$ vs $f_{dc}$ 
for increasing ac amplitudes
of $A =0.25$, 0.275, 0.3, 0.325, 0.35, 0.375, and $0.4$. 
Fig.~6 shows that $V_{x}$ exhibits a series
of steps, most of which have $dV_{x}/df_{dc} = 0$. The $n$th step
has $V_{x} = na \omega/(2 \pi)$. 
For increasing ac amplitudes, more steps can be resolved at higher
values of $f_{dc}$. 
The widths of
some of the steps can be seen to grow with increasing
$A$, while others decrease.  
There are also some regions of drive which do not 

\begin{figure}
\center{
\epsfxsize=3.5in
\epsfbox{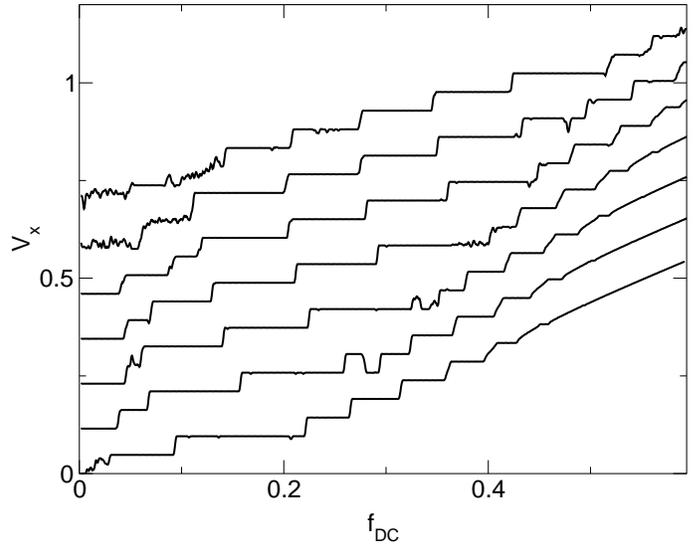}}
\caption{
$V_{x}$ vs $f_{dc}$ for, from bottom to top, $A = 0.25$, 
0.275, 0.3, 0.325, 0.35,
0.375, and $0.4$. The curves have been systematically shifted
in $y$ for clarity.
}
\end{figure}

\noindent
settle onto  
clearly defined steps. These regions become more prominent
for higher values of $A$. For example, in Fig.~6, the 
depinning and the $n=1$ and 2 steps of the
two upper curves $A = 0.375$ and $0.4$
have large fluctuations.  The 
depinning threshold for the bottom curve at $A =0.25$
also shows a similar behavior. These values
of $A$ are close to or at the transition where the
number of potential maxima encircled by the particle orbit 
at $f_{dc} = 0.0$ changes.
For the upper curves, this is from 
one to four maxima, and for the lower curve, it is from zero to one maximum.
By watching animations of the particle orbits, 
we observe that in general the particle
is jumping between the two different orbits on these poorly
defined steps. 
At these values of $A$ the particle orbits are delocalized and 
the depinning threshold is zero, as shown in Fig.~2.  
Another feature of the $V_{x}$ curves 
is that occasionally there are regions where the
velocity jumps down 
in value with increasing $f_{dc}$, such as in the $A=0.275$ curve 
near the transition between the $n=3$ and $n=4$ steps
at $f_{dc} = 0.28$. 
In these cases the particle orbit jumps from the higher $n$ orbit 
back
to the lower state.  In  general these step down effects occur near the 
$n$ to $n+1$ 
transitions. If we repeat the same simulation with 
slightly different initial conditions, similar jumps occur
in the same regions of $f_{dc}$ but are not identical. 
We have previously shown that along the 
flat steps, the particle orbits are stable.   
Along the $n$th step the particle moves a distance $na$ in the
$x$ direction in 
a single period. 

In Fig.~7 we show the particle orbits 
on the integer steps $n = 1, 2, 3$, and $4$
for the system in Fig.~6 at $A = 0.325$.  
For the drives shown, 
$<V_{y}> = 0$.
At zero drive, the orbit has the square shape 
illustrated in Fig.~1(c).
On the $n=1$ step
at $f_{dc} = 0.054$ (Fig.~7(a)), 
the particle circles around a single maximum 
and moves in the $x$ direction by a distance $a$ per period.
For the 

\begin{figure}
\center{
\epsfxsize=3.5in
\epsfbox{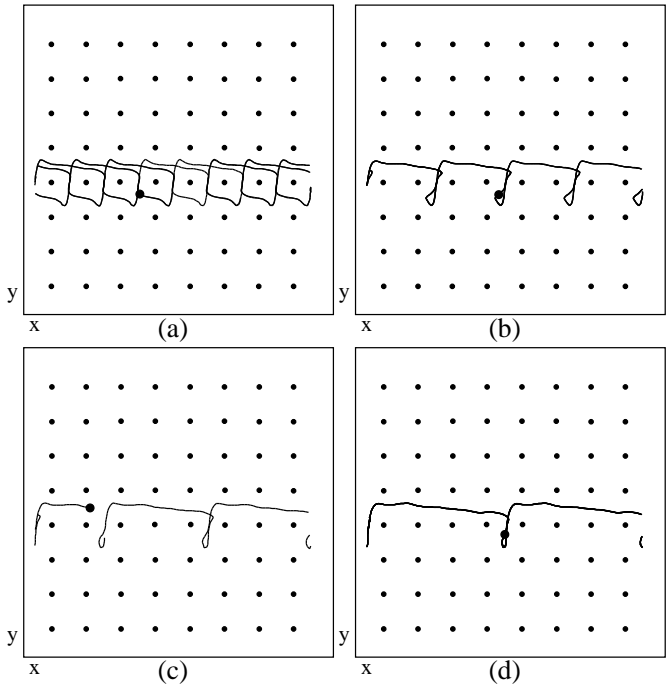}}
\caption{Particle trajectories (black lines), driven
particle (large dot), and
periodic potential maxima (black dots) 
for the first four steps in $V_{x}$ from Fig.~6 for the
$A = 0.325$ curve. (a) $n=1$ step at $f_{dc} = 0.054$.
(b) $n=2$ step at $f_{dc} = 0.1$. (c) $n=3$ step at $f_{dc} = 0.18$.
(d) $n=4$ step at $f_{dc} = 0.24$. 
}
\end{figure}

\noindent
$n=2$ step at $f_{dc} = 0.1$ (Fig.~7(b)), 
the nature of the particle motion changes.
Rather than circling around every second maximum, the
particle moves through a smaller loop that is less than $a$ 
in diameter in every other plaquette so that in one 
period the particle moves a distance $2a$. 
Similar motion occurs on the $n=3$ step
at $f_{dc} = 0.18$ (Fig.~7(c)), 
but the loop occurs every third plaquette. 
Along the $n=4$ step at $f_{dc} = 0.24$ (Fig.~7(d)), the 
particle translates $4a$ in a single period. 

 In Fig.~8 we plot the orbits along the $n=5,$ 7, 8, and 9 steps for the
same system as in Fig.~6 at $A = 0.325$. 
For the $n=5$ step at $f_{dc} = 0.325$, the orbit is essentially the
same as those in steps $n=2$ to $n=4$ from Fig.~7.  
The figure has
a loop in every plaquette since we have shown the particle crossing the
periodic boundary several times, and the particle does not follow
its previous path until it has completed
several passes through the system due to the fact that the orbit
repeats every five plaquettes but the sample has an even number of
plaquettes.
If we chose a system size which is a multiple 
of $5a$ wide, the orbit is repeated during each pass through the system.
Whether the sample size is commensurate with
the orbit shape
does not change any features in $V_{x}$ or $V_{y}$.
On the $n=7$ step at $f_{dc} =0.45$ (Fig.~8(b)), 
the particle
is moving fast enough in the $x$ direction that it 
can no longer loop down into the lower row. 
The orbit still shows a small loop inside the row 
every seventh plaquette. 
For the $n=8$ step shown in Fig.~8(c) at $f_{dc} = 0.49$, 
the particle 

\begin{figure}
\center{
\epsfxsize=3.5in
\epsfbox{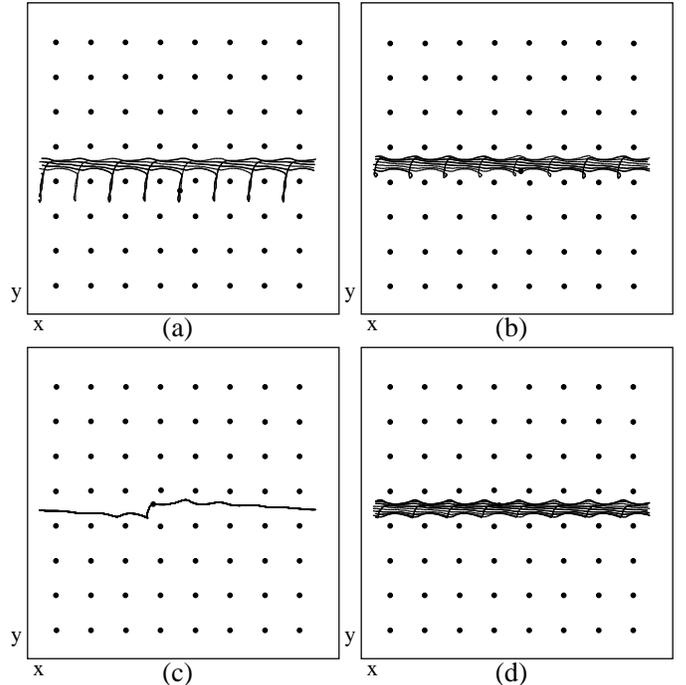}}
\caption{Particle trajectories (black lines), periodic potential maxima
(black dots), and driven particle (large dot) 
for the velocity steps from Fig.~6 at $A = 0.325$.
(a) $n=5$ step at $f_{dc} = 0.325$. (b) $n=7$ step at $f_{dc} = 0.45$.
(c) $n=8$ step at $f_{dc} = 0.49$. (d) $n=9$ step at $f_{dc} = 0.53$. 
}
\end{figure}

\noindent
motion is again contained within
one row and shows a very small loop every $8a$.  The
orbit is commensurate with the sample size so the orbit 
repeats exactly during each pass.  
For the $n=9$ step at $f_{dc} = 0.53$ (Fig.~8(d)), the orbit 
is similar to that seen for the $n=7$ and 
$n=8$ steps, with a small loop every $9a$. 
For much higher $f_{dc}$, the 
particle does not lock to a fixed orbit for this value of $A$.

In addition to the integer steps, we also observe fractional steps 
in the regions between the integer steps.
In general these fractional phases 
are associated with the onset of rectification,
where the average particle velocity is no longer strictly in the
$x$ direction.  

\subsection{Rectifying Phases}
In Fig.~9 we plot 
simultaneously 
$V_{x}$, which increases with
$f_{dc}$, 
and $V_{y}$ vs $f_{dc}$ for a fixed value of $A = 0.325$. 
Here, the steps in $V_{x}$ have height
$a\omega/(2\pi)$, while $V_{y}$ shows {\it non-zero values} centered at the step
transitions in $V_{x}$. The maximum value of $V_{y}$ is $a\omega/(2\pi)$
in the positive direction as seen near the $n=2$ to $n=3$, $n=3$ to $n=4$, 
and $n=4$ to $n=5$ step transitions. 
At the $n=5$ to $n=6$ transition, $V_{y} = -a\omega/(2\pi)$.   
There are also ranges of drive over which
the value of $V_{y}$ is not a multiple 
of $a\omega/(2\pi)$, such as at the $n=0$ to $n=1$, $n=1$ to $n=2$, and 
the $n=5$ to $n=6$ steps.  
Rectification occurs 

\begin{figure}
\center{
\epsfxsize=3.5in
\epsfbox{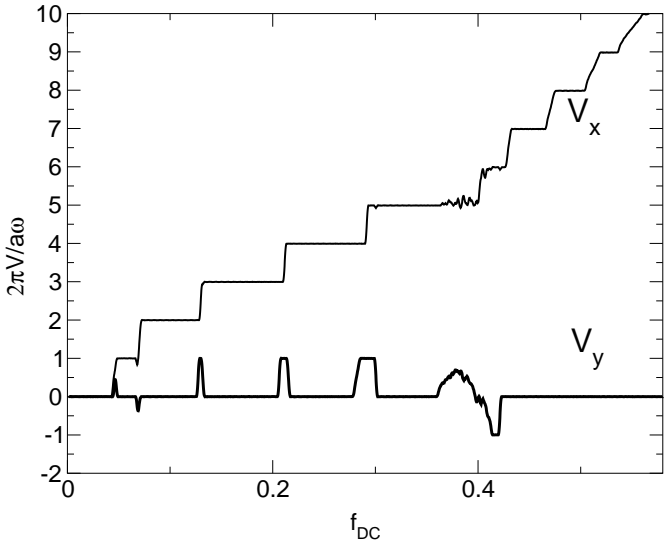}}
\caption{$V_{x}$ (light line) and $V_{y}$ 
(dark line) vs $f_{dc}$ for $A = 0.325$. 
}
\end{figure}

\noindent
everywhere along the $n=6$ step.
No rectification occurs for the $n=7$ step and above, which 
also corresponds to the orbits becoming confined to a single row for
these high $f_{dc}$ values, as illustrated in Fig.~8.

The rectification can be understood by considering the symmetries of the
problem.  The dc drive breaks the reflection symmetry across the $y$-axis,
$R_y$, but preserves $R_x$, reflection across the $x$-axis, as can be
seen by noting that 
the reflection $R_y$ would change the sign of the dc drive (which is
applied in the x-direction) while
the reflection $R_x$ would leave the drive
unchanged.  The dc drive also breaks the combined symmetry $R_xR_y$.
The ac drive breaks both $R_x$ and $R_y$ individually, but preserves the
combined symmetry $R_xR_y$.  Here, 
either the reflection $R_x$ or the reflection $R_y$ reverse the direction
of the ac drive from counter-clockwise to clockwise, but 
the combination $R_xR_y$ leaves
the drive unchanged up to a change in the phase of the drive (corresponding
to a shift in $t$ by half a period).  The combination of the ac 
and dc drives break
all of the symmetries in the problem.

To see the effect of the symmetries, consider first the situation with only
the dc drive, when the system has the symmetry $R_y$.  Then, if the particle
has an orbit with non-zero $V_y$, by symmetry it must also have an orbit
with the opposite $V_y$.  If both such orbits exist, there is a spontaneous
symmetry breaking which can produce a rectification.  
Such a spontaneous symmetry breaking has been observed
in similar systems
\cite{SymB32}.  Similarly,
if we have only the ac drive, the system has the symmetry $R_xR_y$.
Then the existence of an orbit with given
$V_x,V_y$ would imply the existence of an orbit with velocities $-V_x,-V_y$,
and spontaneous symmetry breaking would again be possible.
In the case considered in our simulations, 
since {\it all} symmetries are broken, we can have rectifying orbits
{\it without} any spontaneous symmetry breaking.  We have seen in fact that
the sign of the rectification does not 

\begin{figure}
\center{
\epsfxsize=3.5in
\epsfbox{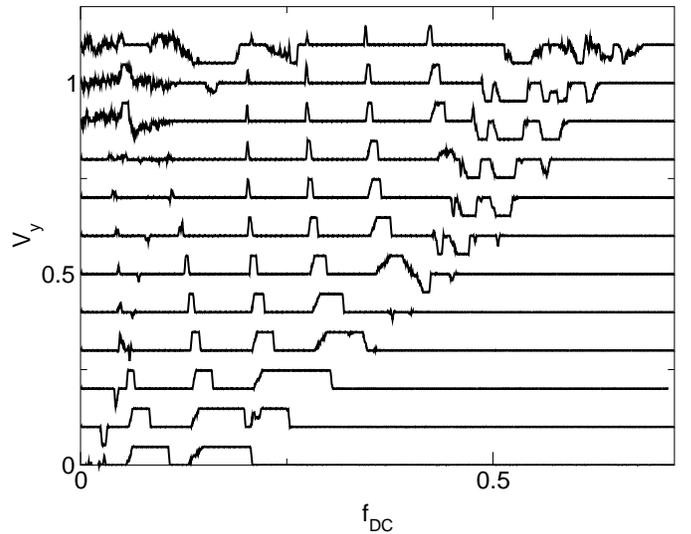}}
\caption{Velocity in the $y$-direction, $V_{y}$, vs $f_{dc}$ for 
$A = 0.25$, 0.262, 0.28, 0.306, 0.31, 0.327, 0.343, 0.356, 0.363,
0.375, 0.38, and $0.4$, from bottom to top. There is a systematic shift
in the $y$-direction added for clarity. 
}
\end{figure}

\noindent
depend on initial conditions, and we
show below that the rectification persists even at non-zero temperature; both
of these are consequences of the fact that the symmetries of the system
are broken by the drives, rather than broken spontaneously.

In Fig.~10 we plot a series of 
$V_{y}$ vs $f_{dc}$ curves for $0.25 \leq A \leq 0.4$,
showing the evolution of the rectifying regions. 
As A increases, the maximum value of $f_{dc}$ at which rectification
is observed also increases,
coinciding with the resolution of more integer steps as shown in Fig.~6.
For $A < 0.32$ (the first five curves from the bottom),
the rectification is predominantly in the 
positive $y$ direction,
while for $A > 0.32$, several phases appear
which rectify in the negative $y$ direction, as seen for 
$A = 0.327$, 0.343, 0.356, and $0.363$. 
For $ A > 0.37$ (the top three curves), 
an increasing number of regions appear
where there is no well defined locking but there is some form 
of rectification. These disordered regions first occur at
low $f_{dc}$ values for $A = 0.375$ and $0.38$, while
for the $A = 0.4$ curve the disordered regions also
appear at higher $f_{dc} = 0.6$. 
The rectification phases shift in position 
with $A$ and the width of the phases grows and then
subsequently shrinks with $A$.
The phases which rectify in the positive direction shift toward lower
$f_{dc}$ with increasing $A$, while the negative rectification phases
shift toward higher $f_{dc}$ as $A$ is increased. 
The shift can be qualitatively understood by considering that 
the particle rotates clockwise.
For the positive rectification regions,
if we consider one cycle starting at a position of
$y=a/2$ and $x=0$ and circling around 
the potential maximum at $(0,0)$,
the particle is moving fastest near $y = +a/2$ when the ac
and dc drives are in the same direction, and
slowest near $y = -a/2$ when 
the ac drive opposes the force from the dc drive.  
If, on the $n$th step, the particle 

\begin{figure}
\center{
\epsfxsize=3.5in
\epsfbox{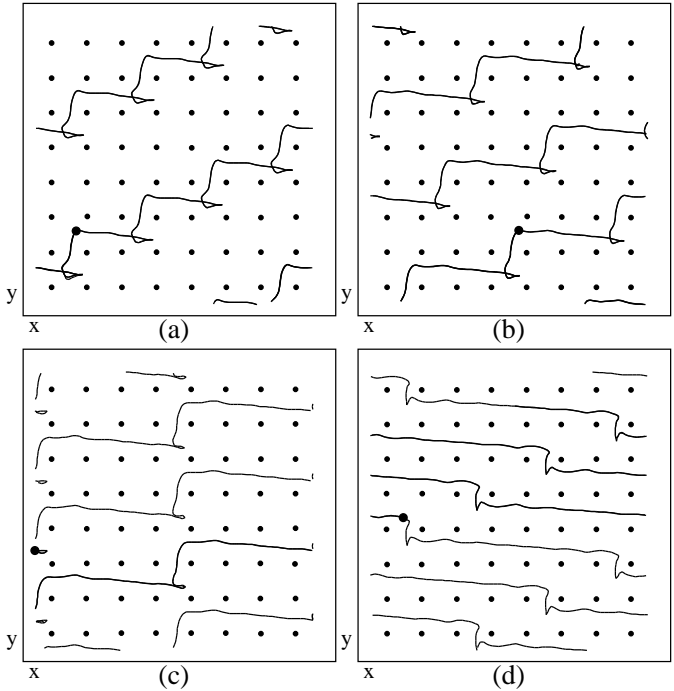}}
\caption{Particle trajectories (black lines), periodic potential maxima
(black dots), and driven particle (large dot), 
showing integer rectifying orbits from the system in Fig.~9 for 
$A = 0.325$ and $f_{dc}=$ (a) 0.1295, (b) 0.21, (c) 0.29, and (d) 0.42. 
}
\end{figure}

\noindent
translates 
a distance $na$
per period, then on the
downward moving portion of the ac orbit, 
the particle interacts strongly with one of the potential maxima. 
If this interaction is too strong, the particle
cannot translate down by one row in the $y$ direction. 
At the same time, the particle
continues to move in the positive $x$ direction.
If, during the downward stroke, the particle moves
a distance close to $na/2$, then on the the upward part of the
ac cycle, the particle does {\it not} 
interact strongly with a potential maximum
and can thus move up in the $y$  direction by a distance $a$. 
As a result, there is a net motion in the $+y$ direction
each cycle.
The positions of the positive rectifying regions shift to 
lower $f_{dc}$ at higher values of $A$ since, 
in the portion of the cycle when the ac and dc forces are in 
the same direction,
a smaller dc drive is required to translate the particle a distance $na$ for 
larger $A$.   
We also note that if the circular ac drive is reversed, the
$V_{y}$ vs $f_{dc}$ curves are flipped and positive rectification
becomes negative rectification.

 In Fig.~11 we show several examples of rectifying phases where 
$V_{y} = a\omega/(2\pi)$ 
for the case of $A = 0.325$ from Fig.~9. 
Figure 11(a) illustrates the first integer rectifying phase for
$f_{dc} = 0.1295$ where there is a transition from the $n=2$ to 
the $n=3$ step. 
In a single cycle, the particle moves $2a$ in the $x$ direction and $a$ in 
the $y$ direction. Additionally, a small loop in the orbit occurs in 
every third plaquette. 
Similar motion occurs on the $n=3$ step at larger $f_{dc}$, where
the particle moves $3a$ 

\begin{figure}
\center{
\epsfxsize=3.5in
\epsfbox{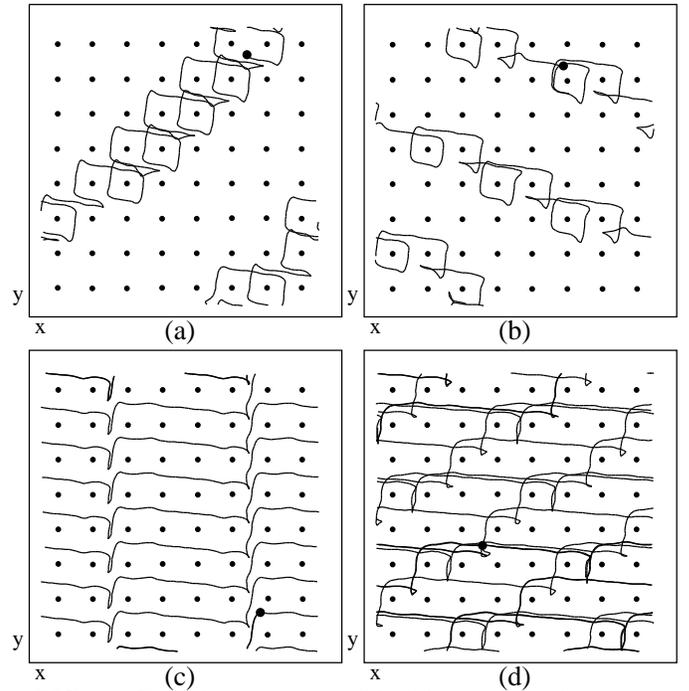}}
\caption{Particle trajectories (black lines), periodic potential maxima
(black dots), and driven particle (large dot) 
showing fractional rectifying orbits from the system in Fig.~9 for 
$A = 0.325$ and $f_{dc} =$ (a) 0.0465, (b) 0.0688, (c) 0.215, and (d) $0.378$. 
}
\end{figure}

\noindent
in the $x$-direction and $a$ in the $y$-direction
during each cycle. 
In Fig.~11(b) we show the orbit for $f_{dc} = 0.21$ for the $n=3$ to
$n=4$ transition,
where the particle moves $3a$ in the $x$-direction in a single cycle.
In Fig.~11(c), near the $n=4$ to $n=5$ transition
for $f_{dc} = 0.29$, the particle moves
$4a$ in the direction of drive during every cycle. The loop feature
that occurs just before the particle translates 
a distance $a$ in the $y$-direction becomes
smaller with increasing step number.  
In Fig.~11(d), we show negative rectification
for $f_{dc} = 0.42$, where the particle
moves $6a$ in the $x$ direction and $-a$ in the $y$-direction in a single 
period.  Here the loop feature seen for the positive rectification
orbits is lost and is replaced
by a kink feature.  

In Fig.~12 we show several examples of fractional rectifying orbits
for $A = 0.325$. Figure 12(a) illustrates the trajectories 
for the rectification at the $n=0$ (pinned) to $n=1$ step transition
at $f_{dc} = 0.0465$. 
In this case, the particle 
moves $2a$ in the $x$-direction and
$a$ in the positive $y$ direction every {\it two} cycles.
To achieve this, the
particle moves $a$ in the $x$-direction and on average $a/2$ in 
the positive $y$ direction in each cycle. 
In Fig.~12(b) we show the 
negative rectification regime near the $n=1$ to $n=2$ transition 
for $f_{dc} = 0.0688$. We find a similar motion as
in Fig.~12(a). Every two cycles, the particle
moves $2a$ in the positive $x$ direction.
The orbit forms a complete loop
around one potential maximum during the first cycle, 
followed by an incomplete
loop in the next cycle, at the end of which the particle has
translated down by $a$. 

\begin{figure}
\center{
\epsfxsize=3.5in
\epsfbox{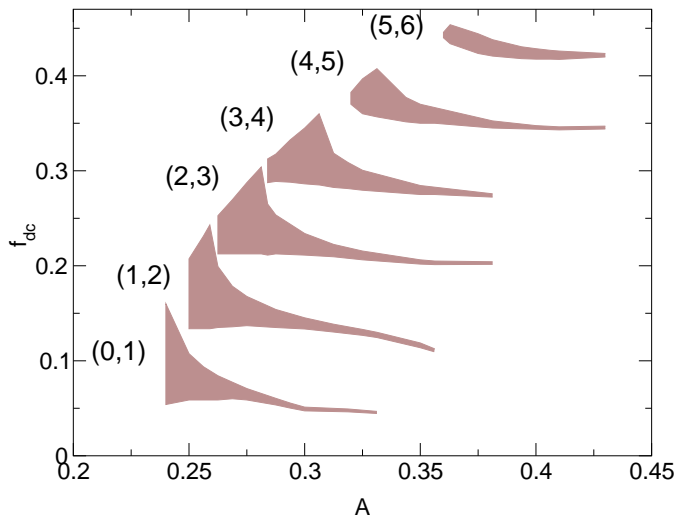}}
\caption{$f_{dc}$ vs $A$, where the shaded regions indicate positive integer
rectifying regions obtained from Fig.~9. 
}
\end{figure}

\noindent
Thus in
a single period, the particle moves $2a$ in the $x$-direction and
an average of $-a/2$ in the $y$-direction. 
We also find that at the onset of the
integer rectifying phases, there can be a small region where the
particle exhibits a fractional rectification. In Fig.~12(c) 
we show one such region that occurs at the end of 
the rectifying phase near the $n=3$ to $n=4$ transition for $f_{dc} = 0.215$. 
Here, in a single period the particle moves $4a$ in the $x$ direction,
while it translates by $a$ in the $y$ direction every other cycle.
During the first cycle, the trajectory dips down and up
but the particle does not translate to the next upper row.
On the following cycle, a cusp forms and the particle moves
up to the next row. There are also several regimes for large $A$
where the particle orbits rectify but do not repeat. For
$A = 0.325$ in Fig.~9, such a regime occurs near
$f_{dc} = 0.378$. In Fig.~12(d) we plot the disordered orbit 
that occurs in this
regime, showing that it has 
a net drift in the $y$-direction. 
We have also examined the rectifying orbits for other values
of $A$ and find that they are similar to those shown
in Fig.~11 and Fig.~12.

\subsection{Rectification Phase Diagram}

In Fig.~13 we plot $f_{dc}$ vs $A$, and 
indicate the occurrence of integer rectification in the positive $y$ direction
by shaded regions. 
The phase diagram has the form of 
a series of tongues, where the 
width of any given rectifying phase decreases for increasing $A$.   
A larger number of rectifying phases appear
at higher dc drive for increasing $A$, 
as seen from the rising envelope which 
begins at $f_{dc} = 0.25$.
The rectifying phases for $f_{dc} > 0.25$ increase in width 
with $A$ over
a small range of $A$ before reaching a maximum width and then decreasing
in width with increasing $A$. 
We do not have the resolution to determine 

\begin{figure}
\center{
\epsfxsize=3.5in
\epsfbox{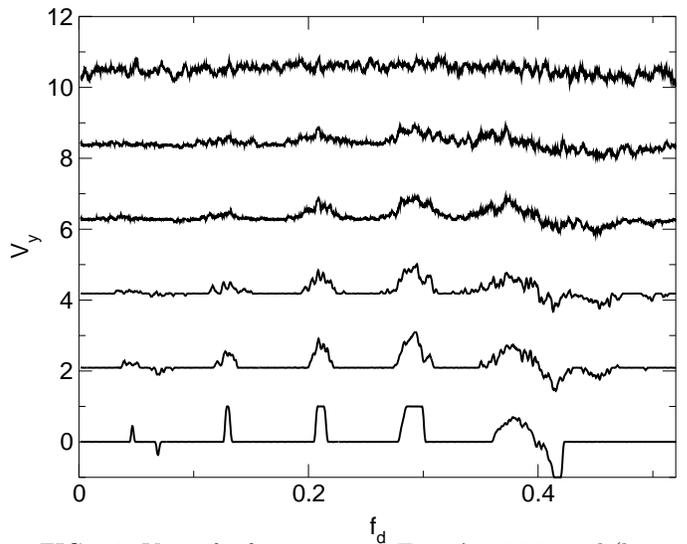}}
\caption{$V_y$ vs $f_{dc}$ for increasing $T$ at 
$A = 0.325$ and (bottom to top) $T = 0.0,$ 0.025,
0.0625, 0.1225, 0.5, and $1.0$.  The curves have
been shifted in the $y$ direction for clarity.
}
\end{figure}

\noindent
whether 
the rectifying phases persist with continually decreasing width 
for higher ac amplitudes, or whether they actually 
terminate. We note that for $A > 0.44$, above the $f_{dc}=0$ transition
from the orbit encircling one to encircling four maxima,
a new set of rectifying phases 
appear at low $f_{dc}$, not shown in the figure. 

We note that it is difficult to plot a phase diagram for the
regions of negative rectification that occur for $A > 0.35$. 
Here the locking steps become hard to define due to the disordered regions
where $V_{y}$ does not settle down to a single value.
In general, the negative rectification
regimes show similar features to the positive rectification regions,
with the width of the rectifying regime growing to a maximum value
and then decreasing for
increasing $A$. 
The phases also shift to higher $f_{dc}$ for increasing $A$. 

\section{Effects of Disorder}

\subsection{Thermal Disorder}
In many experimentally realizable systems such  as
colloids or biomolecules, thermal effects or Brownian motion are
relevant. To model thermal effects, we add a noise term $f^T$ 
to the equation of motion, with the properties
$<f^{T}(t)> = 0.0$ and
$<f^{T}(t)f^{T}(t^{\prime})> = 2\eta k_{B}\delta(t- t^{\prime})$. We have 
performed a series of simulations for $A = 0.3$ for different values of 
temperature. In Fig.~14 we show $V_{y}$ vs $f_{dc}$ for 
$A = 0.325$ for increasing temperature. 
For low $T$ there are still regions of $V_{y} \approx 0$
within our resolution. 
The particle orbit at low temperatures shows
only small perturbations, so the behavior is thermally activated
and $V_{y}$ is not strictly zero but instead is 
exponentially small. For higher
$T$, the orbits are strongly perturbed, and the maximum value of $V_{y}$
decreases while the width
of the 

\begin{figure}
\center{
\epsfxsize=3.5in
\epsfbox{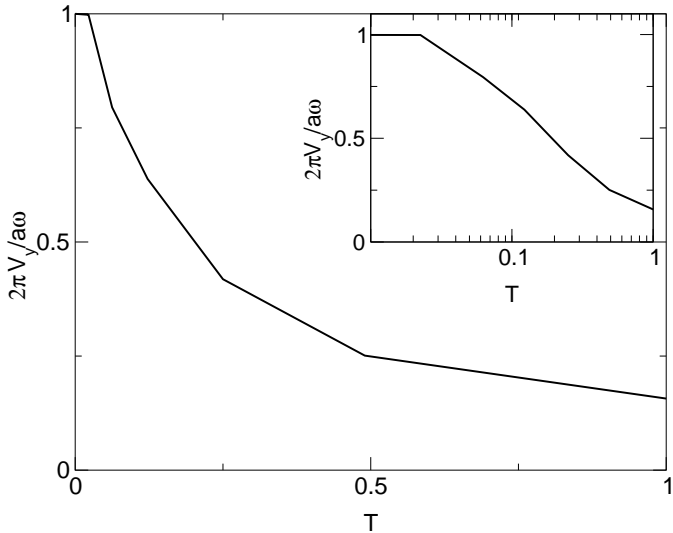}}
\caption{$2\pi V_{y}/a\omega$ vs $T$ for $A = 0.325$ and fixed $f_{dc} = 0.29$.
The inset is a log-linear plot of the main figure. 
}
\end{figure}

\noindent
rectifying regimes increases. For the highest temperatures, the 
particle diffuses rapidly; however, some rectification
still occurs. For the top curve ($T=1$) there is only 
a slight positive rectification 
for $f_{dc} < 0.4$.
As $T$ increases further the rectification is
gradually completely lost. 
In Fig.~15 we plot the average $V_{y}$ for the positive rectification
region at $f_{dc} = 0.29$ for the bottom curve in Fig.~14 at $T = 0.0$.
We define the temperature scale such that at
$T = 1.0$, a single particle
with $f_{dc} = 0.0$ begins to diffuse, indicating that
$T=1$ is the melting temperature 
for our parameters. In our system we have only one colloid, so
thermal activation occurs at a much lower temperature than for 
a collection of interacting colloids. 
Fig.~15
shows a decay of $V_y$ with increasing $T$, 
as illustrated more clearly in the inset.
A reliable fit can be applied,
giving $2\pi V_{y}(T)/a\omega = (1 - \exp(-B/T))$, which indicates  
thermal activation. The temperature scale can be changed by increasing
the depth of the periodic wells as well as by the addition of other
particles, which alters the effects of fluctuations. 

\subsection{Particle Interactions} 

In many experimentally relevant systems it is likely that multiple particles
would be moving through the array at the same time. If the density 
of the particles is sufficiently high, particle-particle interactions or
scattering become relevant. 
For low fillings where the
particles are still far apart, 
the $V_{y}$ curves are only weakly perturbed. As the
filling fraction is increased, the $V_{y}$ curves exhibit some disordering. 
There are certain higher filling fractions such as $1/16$
where the $V_{y}$ curve is virtually identical to the single particle case.
This is due to a commensuration effect. 
For particles with long range interactions such as vortices in superconductors
or colloids which are weakly screened, commensuration effects 
will occur for different particle densities. 
For fillings such as $1/2$ where
there is one mobile particle for every other plaquette, the particles
form an ordered arrangement. In the case
of half filling, the particles form a checkerboard state. Similar ordered 
states
occur at fillings of $1/16$, 1/8, 1/5, and $1/4$. 
Since the 
arrangements are symmetrical at these fillings, the 
interactions effectively cancel and the system shows the same behavior
as the single particle case.  At incommensurate 
fillings where ordered particle arrangements 
cannot be formed, the particle-particle
interactions become relevant. 

\section{Simple Model System} 

Let us return to the quantization of the velocity discussed above,
and consider both the stable plateaus and the intermittent
transitions between plateaus 
using general properties of nonlinear maps.
Define
a map $(x,y)\rightarrow (x'+n_x a,y'+n_y a)$, from the position of the particle
at the start of a period to that at the end, where we may restrict to
$0\leq x,y,x',y'\leq a$, with $n_x,n_y$ integer.  Here, we have taken
$x,y,x',y'$ to indicate positions of the particle within a unit cell,
while $n_x,n_y$ indicate which unit cell the particle occupies after one
orbit.  We have fixed the unit cell of the particle at the start of the
orbit; by translation symmetry, it does not matter which cell this is.
If there is a stable
fixed point, $(x,y)=(x',y')$, then the particle translates by
$(n_x a,n_y a)$ in time $2\pi\omega^{-1}$ and so has average velocity $V_x,V_y$
quantized in multiples of $a \omega/(2\pi)$, as found above.
If the $q-th$ power of the map has a stable fixed point,
there are instead steps of fractional heights $(p/q) a \omega$.

As $f_{dc}$ increases, the periodic orbit becomes unstable,
and
a different periodic orbit with larger $V_x$
appears.  This new orbit will be the next stable periodic orbit
at higher drive.  The
transition to the new orbit can occur in one of three ways.
{\it (1)}  If both periodic orbits are stable simultaneously,
the particle velocity will depend on the initial conditions
in the transition regime.   This was not observed.
{\it (2)}  The second periodic orbit
could become stable at the same time that the first orbit becomes
unstable.  This behavior, which gives rise to infinitely sharp jumps
in $V_x$,
is not generic and hence not expected.
{\it (3)}  There can be a finite range of drive containing
no stable periodic orbits.  Over this range, the average velocity is
not quantized.  If, however, some orbits are close to stable, the
particle will spend long times in these orbits, giving rise to intermittent
behavior.  This last behavior is consistent with what we observe.

The stability of the orbits is shown by the fact that in the middle of
the plateaus the system is not sensitive to initial conditions, and by the
exponentially small change in velocity at non-zero temperature.  Outside
the plateaus, the change in average velocity most likely will 

\begin{figure}
\center{
\epsfxsize=3.5in
\epsfbox{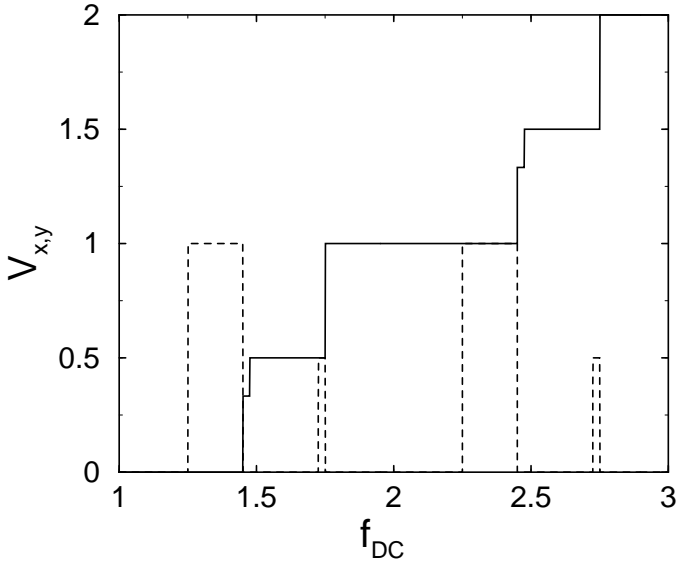}}
\caption{Time averaged $V_x$ (solid line) and $V_y$ (dashed line)
as a function of $f_{dc}$ in toy model I.
}
\label{fig:Matt}
\end{figure}

\noindent
not show
a thermally activated form.  We were not able to accurately measure the
velocity in these reasons well enough to determine the temperature
dependence of the velocity.

We now turn to a specific toy model illustrating some of these
ideas.  Consider a particle in a lattice of repulsive sites
with $a=1$, where the potential minima between repulsive sites are
at integer $x$ and $y$ values.  The $y$ position of the
particle is constrained
to take only integer values, but the $x$ position can
be any real value.  To model the translation of the particle
through the lattice,
we separate the $x$ and $y$ motion,
so that the particle moves first ({\it i.}) right
at velocity $v_r$, then ({\it ii.}) down,
then ({\it iii.}) left at velocity $v_l$, 
then ({\it iv.}) up.  In some cases, steps ({\it ii.}) and ({\it iv.}) of the
cycle may not produce a change in $y$ due to the constraint that
$y$ can be only integer valued; this corresponds to the confinement
of the particle to a single row in the physical system.
We consider two 
slightly different
versions of the model.  The first version, I, follows the following
sequence of transitions:
I({\it i.}): We apply the rule $x \rightarrow x+v_r$.
I({\it ii.}):  If $x$ is within $0.25$ of an integer, $x$
is set to that integer and $y$ is decremented by one.
I({\it iii.}): Apply $x \rightarrow x-v_l$.
I({\it iv.}): As in ({\it ii.}) except $y$ is incremented by one.
Here $v_r$ and $v_l$ are the velocity of the particle in the rightward
and leftward parts of the cycle, respectively.
In steps I({\it ii.}) and I({\it iv.}), the particle will {\it only} move
to a new $y$ position if it reaches the minima between sites
at the correct phase of the driving period, when transverse motion is
possible.  In this case, the particle slips into the next row and
the $x$ coordinate of the particle is set to an integer value.
In Fig.~\ref{fig:Matt} we show the time-averaged velocities $V_x$ and $V_y$
obtained with model I for fixed $v_l=1.2$ and increasing
$v_r$, representing increasing $f_{dc}$.  It is clear that
this simple model produces plateaus in $V_x$ as well as ratcheting 

\begin{figure}
\center{
\epsfxsize=3.5in
\epsfbox{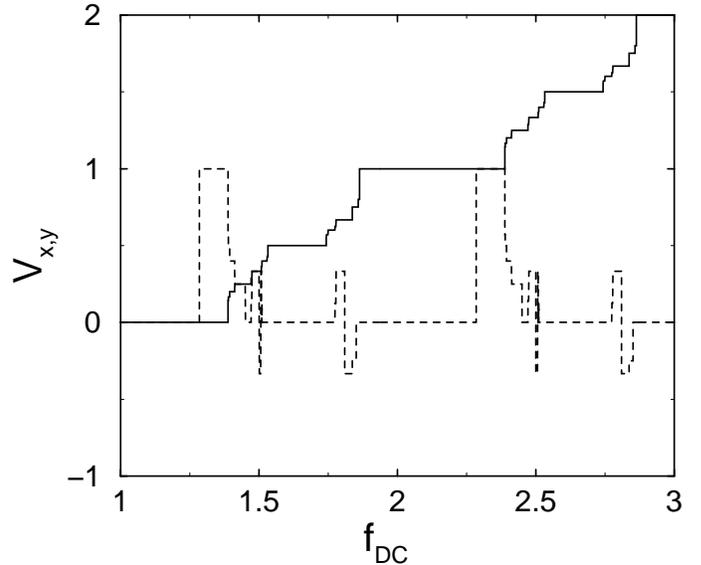}}
\caption{Time averaged $V_x$ (solid line) and $V_y$ (dashed line)
as a function of $f_{dc}$ in toy model II. 
}
\label{fig:Matt2}
\end{figure}

\noindent
behavior
in $V_y$.
The sharp jumps in the velocity values are
due to the discontinuity of the map function,
as in case {\it (2)} above;
for smoother map functions, these
jumps
acquire a small but finite width.
More complicated maps
produce
richer behavior, including occasional regimes of
negative $V_y$.  To show this, we consider a slightly different
version of the model.  In this model II, if the $x$-position of the
particle is within a
distance of $0.25$ of an integer in step II$(ii.)$ or II$(iv.)$, 
the $x$-position
is set equal to $0.25$ times the present $x$-position plus $0.75$ times the
given integer.  This smoother map allows a much richer behavior; if the
particle is within $0.25$ of an integer in the
model I, the final position of the map does not depend on exactly where
the particle position is near the given integer, while 
in model II the final position
{\it does} 
depend on the precise particle position.  We show the average velocities 
for this model in Fig.~\ref{fig:Matt2}.

The ratcheting behavior in both the model and simulations
occurs near transitions in $V_x$ when the number of pinning centers the
particle passes in one period changes, making it possible for the particle
to interact asymmetrically with the pinning sites.  For a clockwise orbit,
the particle moves rapidly on the upper portion of the orbit, and is
likely to scatter off the pinning site below
when the orbit does not quite match $na$.  On the lower
part of the orbit, however, the particle is moving more slowly, and is
likely to slip between the pinning sites above in spite of a small
mismatch.
The particle thus tends to ratchet in the positive $y$ direction.
If the dc drive is reversed,
downward motion should be preferred, as we observe.

Finally, we note that much of this behavior is specific to two or more
dimensions, or to systems in one dimension which are not overdamped so
that both position and momentum are independent degrees of freedom.  Consider
 a map $x\rightarrow x'$, subject to $x+a \rightarrow x'+a$ and
${\rm d}x'/{\rm d}x\geq 0$, true for overdamped motion in one
dimension.  It can then be shown that it is not possible to have
periodic orbits with different values of the velocity as follows: 
Suppose there
were two such orbits.  Then, let initial conditions for the two orbits
be $x_1$ and $x_2$, and let the orbit starting
at $x_1$ have a greater velocity than that starting at $x_2$.
Suppose that $x_1<x_2$ (this can always be accomplished
by translating $x_1$ by some number of unit cells); then, after some
number of mappings, $f(f(...(x_1)...))>f(f(...(x_2)...))$, which violates
the assumption that
${\rm d}x'/{\rm d}x\geq 0$.  Thus, all periodic orbits must have the same
value of the velocity.
Systems with Shapiro steps
do not exhibit jumps.

\section{Discussion}

We now consider physical systems where the phase locking and the
rectification might be realized. 
One possibility is colloids moving through a
periodic 2D array. The array can constructed from a substrate of 
hard obstacles or more smoothly varying objects. If the colloids are charged
they can be driven with dc and ac electric fields. 
The most promising approach would
be to use periodic arrays of optical traps 
\cite{Bechinger33,Dufresne34,Brunner35,Kordaa36} 
or dynamical optical trap arrays \cite{Grier11,GrierN13}.
In this case colloids can be trapped at
individual spots of laser light. 
Once the array is filled, additional colloids move 
through the periodic potential created by the pinned colloids.
One advantage of the light arrays is that the array itself can be 
rotated dynamically to 
mimic ac driving when only an external dc force is applied. 
Recent experiments \cite{Korda12} have demonstrated
the flow of colloids in 2D through periodic optical trap arrays.
Another system would be vortices in superconductors driven with 
ac and dc currents. Periodic arrays of pinning sites can be lithographically
constructed \cite{Moshchalkov37,Schuller38}. 
With the application of a magnetic field, 
flux enters in the form of quantized vortices. If the 
pinning sites are small enough that only a single
vortex can be trapped at each site, then beyond the first matching field 
additional vortices will sit in the interstitial regions. 
It should also be possible to create
arrays of anti-pinning sites, such as with an array of
magnetic dots that are magnetized in the same direction as the
applied field and create fixed vortices.
Additional vortices created by the external field
will move in the interstitial regions between the fixed vortices.
These effects may also occur for fluxons in
2D  Josephson junction arrays
driven with a dc drive and a circular ac drive. In this  
case the fluxon can be viewed as a classical particle moving over a 
2D periodic potential.

\section{Summary}

To summarize, we have investigated the dynamics of
overdamped particles moving 
in a 2D symmetrical periodic array where the particles are driven with
a dc drive in the longitudinal direction and a circular ac drive. 
For small ac drives, we observe phase locking 
in the form of steps in the longitudinal
velocity when the frequency of the ac drive matches with the
frequency of the internally generated ac velocity component. 
For ac drives large enough that the particle can encircle at least 
one potential maximum at zero dc drive, we find
phase locking steps in the longitudinal velocity for increasing dc drive.
Additionally, in this regime we observe a non-zero transverse velocity
in either the positive or negative direction in spite of the fact that 
there is no
dc transverse applied drive. This rectification in the transverse 
direction arises due to the symmetry breaking caused
by the circular ac drive.
We propose and examine a more simplified model
of the system that reproduces many of these
features that we observe. The results of the simple model suggest that the 
phase locking and rectification phenomena described here
are a general feature of a wide class of similar systems.
We show that stable particle orbits occur along the 
longitudinal and transverse steps, while
more chaotic orbits appear in non-step regions.
Finally, we show that thermal disorder and incommensuration 
can smear or reduce the step size,
but regions of rectification still occur.
Our results should be testable for dc and driven vortex motion and colloids
through 2D periodic arrays.

Acknowledgments---We thank
C. Bechinger, M. Chertkov, 
D.G. Grier, P.T. Korda,
and Z. Toroczkai 
for useful discussions. 
This work was supported by the US DOE under Contract No. W-7405-ENG-36.

\end{document}